\documentclass{article}

\usepackage{amssymb}
\usepackage{amsthm}
\usepackage{amsmath}

\usepackage{graphicx}

\newtheorem{theorem}{Theorem}[section]

\newtheorem{conj}[theorem]{Conjecture}
\newtheorem{remark}[theorem]{Remark}

\begin{document}

\title{Gradient catastrophe and flutter in vortex filament dynamics}
\author{
B.G.Konopelchenko $^1$ and G.Ortenzi $^{2}$ \footnote{Corresponding author. E-mail: giovanni.ortenzi@unimib.it,  Phone +39(0)264485765 Fax:+39(0)264485705.  }\\
$^1$ {\footnotesize Dipartimento di Fisica, Universit\`{a} del Salento 
and INFN, Sezione di Lecce, 73100 Lecce, Italy} \\
 $^2$ {\footnotesize Dipartimento di Matematica Pura ed Applicazioni, 
Universit\`{a} di Milano Bicocca, 20125 Milano, Italy}
}\maketitle
\begin{abstract}
Gradient catastrophe and flutter instability in the motion of vortex filament within the localized induction approximation are analyzed. It is shown that the origin of this phenomenon is in the gradient catastrophe for the dispersionless Da Rios system which describes motion of filament with slow varying curvature and torsion. Geometrically this catastrophe manifests as a rapid oscillation of a filament curve in a point that resembles the flutter of airfoils. Analytically it is the elliptic umbilic singularity in the terminology of the catastrophe theory. It is demonstrated that its double scaling regularization is governed by the Painlev\'e-I equation. 
\end{abstract}
\section{Introduction}
Motion of a thin vortex filament in an incompressible inviscid fluid in an infinite three-dimensional domain within the localized induction approximation (LIA) is governed by the simple  equation for the induced velocity $\vec{v}$
\begin{equation}
\label{LIA-X}
 \vec{X}_t(s,t)\equiv \vec{v}(s,t)= \vec{X}_s \wedge \vec{X}_{ss}=K(s,t) \vec{b} 
\end{equation}
which implies the following intrinsic equations for the curvature $K$ and torsion $\tau$
\begin{equation}
\label{LIA-Eqn}
\begin{split}
 K_t&=-2K_s \tau -K \tau_s, \\
\tau_t &= KK_s -2\tau\tau_s + \left( \frac{K_{ss}}{K}\right)_s.
\end{split}
\end{equation}
Here $\vec{X}(s,t)$ is a position vector for a point on the curve representing the filament, $t$ is time, $s$ is the arclenght parameter, subscripts indicate the differentiation with respect to the indicated variables, and $\vec{b}$ is the binormal. \par
 Equations (\ref{LIA-X}) and (\ref{LIA-Eqn}) have been derived by L. S. Da Rios in 1906 {\cite{DR}} and have been rediscovered sixty years later in \cite{AH,Bet}.
The detailed history of Da Rios (DR) system  is presented in \cite{Ric}
For the localized induction approximation see e.g. \cite{Bat,saf}.\par
In 1972 Hasimoto has observed  \cite{Has} that equation (\ref{LIA-X}) is equivalent to the focusing nonlinear Schr\"{o}dinger equation (NLS)
\begin{equation}
 \label{NLS}
i\psi_t +\psi_{ss}+\frac{1}{2} |\psi|^2\psi=0
\end{equation}
 via the transformation 
\begin{equation}
\label{Has-trans}
 \psi(s,t)=K(s,t) \exp \left(i \int ^s \tau(s',t) ds'\right).
\end{equation}
Only one year before Zakharov and Shabat \cite{ZS} discovered that NLS equation (\ref{NLS}) is integrable by the inverse scattering transform (IST) method. Hasimoto's result has demonstrated that the whole powerful machinery of the IST method (solitons, infinite sets of integrals of motion, symmetries, etc.) is applicable to the vortex filament dynamics and has lead to the explosion of interest to the vortex filament dynamics. Since that time various aspects of this dynamics have been analysed by different methods (see e.g. \cite{Lamb}--\cite{SC}).\par
 Stability of vortex filament motion and formation of singularities are two important problems partially addressed during this period (see e.g. \cite{FM,KM,GRV,BV}). 
Modulation instability of the focusing NLS  equation (\ref{NLS}) (see e.g. \cite{FL,EGKK,Agr}) definitely is a key element in such an analysis. Recently it was shown that the dispersionless limit of the NLS equation (\ref{NLS}) is quite relevant to the study of the modulation instability \cite{MK}--\cite{BT2}. In this limit the focusing NLS equation is the elliptic system of quasilinear PDEs. The Cauchy problem is ill-posed for this system and solutions exhibit the gradient catastrophe type behavior at finite time. Behavior of solutions of the dispersionless NLS equation near the point of gradient catastrophe and beyond it has been studied in \cite{DGK,BT,BT2} too. \par
In the present paper we study the geometrical and analytical implications of such an analysis for the vortex filament dynamics. For this purpose it is convenient to deal with Da Rios system (\ref{LIA-Eqn}) directly. We first consider a dispersionless (dDR) system which describes motion of vortex filament with curvature and torsion slowly varying along the filament and in the motion. It is the system of quasilinear equations of elliptic type which in terms of the complex-valued Riemann invariant $\beta=-\tau+iK$ and slow variables $x=\epsilon s$, $y=\epsilon t$, $\epsilon \ll 1$ is of the form
\begin{equation}
\label{betay}
 \beta_y=\frac{1}{2} \left( 3\beta + \overline{\beta} \right) \beta_x.
\end{equation}
 This system well approximates DR system (\ref{LIA-Eqn}) until the derivatives $\beta_x$, $\beta_y$ are not large. Hodograph equations for equation (\ref{betay}) describe critical points of the function $W$ which obey the Euler-Poisson-Darboux equation $E\left(\frac{1}{2},\frac{1}{2}\right)$. \par
Gradient catastrophe for the dDR system happens in a point $(x_0,y_0)$ for given initial data. At this point $\beta_x$ and $\beta_y$ (or $K_x$, $K_y$, $\tau_x$, $\tau_y$ ) explode.
The acceleration $\vec{a}=\vec{v}_t$ explodes too. So the filament becomes fast oscillating near the point $x_0$ at the ``time '' $y_0$. Numerical analysis (borrowed from \cite{CMM}) shows that such oscillations begin to expand along the filament. Intrinsically, the filament begins to oscillate around the rectifying plane near the point $x_0$. So, the gradient catastrophe for the system (\ref{betay})  gives rise to fast oscillation which can be referred as filament flutter  by analogy with airfoil flutter (see e.g. \cite{MilT}). \par
Analysis of behavior near the point $x_0,y_0$ of gradient catastrophe shows that $\beta=\beta_0+\epsilon \beta^*$, $x=x_0+\epsilon x^*$, $y=y_0+\epsilon y^*$, $\epsilon \ll 1$, and
\begin{equation}
 W=W_0+\epsilon^3\left(x^*(f U + g V)+\frac{1}{3}U^3-UV^2\right)
\end{equation}
where $U+iV \propto \beta$ and $f$, $g$ are some constants. Thus the function $W$ exhibits an elliptic umbilic singularity behavior in the terminology of Thom \cite{Thom}. Regularization of this singularity is one of the issues of this paper.\par
 At the point of gradient catastrophe the approximation (\ref{betay}) becomes invalid and should be substituted by the full DR system (\ref{LIA-Eqn}). First order  corrections can be obtained by the double-scaling technique together with appropriate modification of the function $W(\beta,\overline{\beta})$ to a functional $W^*$ such that the equations of critical points for $W$ are substituted by the Euler-Lagrange equation for $W^*$. The resulting equation for small corrections is equivalent to the Painlev\'e-I (P-I) equation 
\begin{equation}
 \label{P-I}
\Omega_{\xi\xi}=6\Omega^2-\xi.
\end{equation}
 The result of the paper \cite{DGK} allows us to conjecture that any generic solution of the gradient catastrophe behaves as $\beta=\beta_0+\epsilon \beta^*$ where the correction $\beta^*$ is described by the tritronque\'e solution of the P-I equation. \par
The paper is organized as follows.
Dispersionless DR system, its different forms, hodograph equation and other its properties are discussed in section 2. Gradient catastrophe for the dDR system is analysed in section 3. Geometrical implications of gradient catastrophe and flutter of filament are considered in section 4.  In section  it is shown that near to the singular point filament exhibits the elliptic umbilic catastrophe behavior. Regularization of this singularity via the Painlev\`e-I equation is presented in section 6. 
%%%%%%%%%%%%%%%%%%%%%%%%%%%%%%%%%%%%%%%%%%%%%%%%%%%%%%%%%%%%%%%%%%%%%%%%%%%%%%%%%%%%%%%%%%%%%%%%%%%%%%%%%%%%%%%%%%%%%%%
\section{Dispersionless Da Rios system}
In order to describe dispersionless (or quasiclassical) limit of DR system one, in a standard manner,  introduces slow variables  $x=\epsilon s$, $y=\epsilon t$ with $\epsilon \ll 1$  and assume that curvature and torsion are smooth functions of slow variables $K=K(x,y)$ and $\tau=\tau(x,y)$. Under these assumptions equation (\ref{LIA-Eqn}) take the form 
\begin{equation}
\label{LIA-Eqn-e}
\begin{split}
 K_y&=-2K_x \tau -K \tau_x, \\
\tau_y &= KK_x -2\tau\tau_x + \epsilon^2 \left( \frac{K_{xx}}{K}\right)_x.
\end{split}
\end{equation}
Thus in the limit $\epsilon \to 0$ one has the dDR system 
\begin{equation}
 \label{dLIA-Eqn}
\begin{split}
 K_y&=-2K_x \tau -K \tau_x, \\
\tau_y &= KK_x -2\tau\tau_x.
\end{split}
\end{equation}
Analogously to the dispersionless NLS equation \cite{KMcM}--\cite{BT} the solutions of the dDR system (\ref{dLIA-Eqn}) well approximate solutions of the DR system in points were $K_x$ and $\tau_x$ are finite. As any two-component system it is linearizable by hodograph transformation $(x,y) \leftrightarrow (K,\tau)$. The characteristic velocities for dDR system are complex $\lambda=-2\tau + iK$, Riemann invariant $\beta=-\tau + iK$ and in terms of $\beta$ the dDR system is of the form (\ref{dLIA-Eqn}). We note that solutions of the dNLS system discussed in \cite{DGK} and those of dDR system (\ref{dLIA-Eqn}) are connected by the simple relations $u=K^2$, $v=2\tau$. For the geometrical consideration the system (\ref{dLIA-Eqn}) is more convenient.\par
The dDR system is known to be integrable similar to its hyperbolic version, i.e. the 1-layer Benney system  \cite{Zak}. It has an infinite set of symmetries and integrals of motion. One of the form of the dDR hierarchy is given by the set of equations (see e.g. \cite{KK}) 
\begin{equation}
\label{dLIA-hie}
 p_{y_n}=\left(\left( \frac{z^n}{p}\right)_+p\right)_x, \qquad n=1,2,3,\dots
\end{equation}
 where $p=z+p_1(x,y)z^{-1}+p_3(x,y)z^{-3}+p_5(x,y)z^{-5}+\dots$ is a formal Laurent series defined by the equations
\begin{equation}
 \label{curve}
p^2=(z-\beta)(z-\overline{\beta})=(z+\tau)^2+K^2,
\end{equation}
$y_n$ are time variables and $f_{+}$ denotes the polynomial part of $f$. The dDR system  (\ref{dLIA-Eqn}) is the first flow of the hierarchy (\ref{dLIA-hie}) at $n=1$. All $p_{2k+1}(x,y)$ are densities of integrals of motion for the dDR system. They are the dispersionless limit of the densities of integrals of motion for the full DR system (\ref{LIA-Eqn}) found in \cite{LP,Ric2}.\par
Solutions of the dDR system can be calculated via the standard hodograph equation. It was shown in \cite{KMM,KMM2} that these hodograph equations are, in fact, the equations
\begin{equation}
 \label{Eul-Lag}
W_{\beta}=0,\qquad W_{\overline{\beta}}=0,
\end{equation}
 which define the critical points of the function of the form
\begin{equation}
 \label{W}
W=\frac{x}{2}(\beta+\overline{\beta})+\frac{y}{8}(3\beta^2+2\overline{\beta}\beta+3\overline{\beta}^2)+\widetilde{W}(\beta,\overline{\beta})
\end{equation}
  where the function $\widetilde{W}(\beta,\overline{\beta})=\widetilde{W}(\overline{\beta},\beta)$ is defined by the initial data for $\beta$ and it is such that $W$ obeys the Euler-Poisson-Darboux equation $E\left(\frac{1}{2},\frac{1}{2}\right)$, i.e. 
\begin{equation}
 \label{EPDe}
2(\beta-\overline{\beta})W_{\beta \overline{\beta}}=W_{\beta}-W_{\overline{\beta}}.
\end{equation}
Since the function $W$ is real valued, the second equation of (\ref{Eul-Lag}) is the complex conjugated to the first one.
Hodograph equation is 
\begin{equation}
\label{hodoW}
 W_{\beta}=\frac{x}{2}+\frac{y}{4}(3\beta+\overline{\beta})+{\widetilde{W}}_{\beta}=0.
\end{equation}
In terms of $K$ and $\tau$ the function $W$ is 
\begin{equation}
 \label{W-tK}
W=-x \tau + y \left( \tau^2-\frac{1}{2}K^2\right)+\widetilde{W}(\tau,K),
\end{equation}
the hodograph equations (\ref{hodoW}) are given by
\begin{equation}
\begin{split}
 W_K=-yK+{\widetilde{W}}_K=0 \\
W_\tau = -x+2y\tau +{\widetilde{W}}_{\tau}=0,
\end{split}
\end{equation}
while the Euler-Poisson-Darboux equation is of the form
\begin{equation}
 \label{Beltrami}
K(W_{KK}+W_{\tau \tau})+W_K=0.
\end{equation}
It is the axisymmetric three-dimensional Laplace equation studied by Beltrami \cite{Bel} (see also \cite{Wei}). Here $K$ and $\tau$ play the role of the radial and axial coordinate respectively in the cylindrical system of coordinates .\par
For the dDR hierarchy (\ref{dLIA-hie}) the function $W$ has the form
\begin{equation}
 W=\sum_{l=0}^{\infty}  y_l \oint_{\gamma} \frac{dz}{2\pi i} \frac{z^{l+1}}{p(z)}
\end{equation}
where $\gamma$ denotes a small circle around infinity. Euler-Poisson-Darboux equations (\ref{EPDe}) is quite useful in the study of singular sector of the 1-layer Benney and dDR hierarchies \cite{KMM,KMM2} 
%%%%%%%%%%%%%%%%%%%%%%%%%%%%%%%%%%%%%%%%%%%%%%%%%%%%%%%%%%%%%%%%%%%%%%%%%%%%%%%%%%%%%%%%%%%%%%%%%%%%%%%%%%%%%%%%%%%%%%%
\section{Gradient catastrophe}
Hodograph equations (\ref{Eul-Lag}) provide us with unique solution of the dDR system if the standard condition
\begin{equation}
 \label{Delta-cond}
\Delta\equiv \det \left( \begin{array}{cc}
                          W_{\beta\beta} & W_{\beta \overline{\beta}} \\
                          W_{\overline{\beta} \beta} & W_{\overline{\beta} \overline{\beta}} 
                         \end{array}
  \right) \neq 0
\end{equation}
is satisfied. Due to the Euler-Poisson-Darboux equations (\ref{EPDe}) on the solutions of the dDR system with 
$K \neq 0$ the function $W$ obeys the equation $W_{\beta \overline{\beta}}=0$.  Hence
\begin{equation}
\Delta= |W_{\beta \beta}|^2
\end{equation}
and, consequently, the hodograph equations (\ref{hodoW}) are uniquely solvable if $W_{\beta \beta}\neq 0$. In this situation the derivatives $\beta_x$ and $\overline{\beta}_x$ are bounded together with $\beta$. Indeed, differentiating equation  (\ref{hodoW}) with respect to $x$, one gets the system 
 \begin{equation}
  \left( \begin{array}{cc}
                          W_{\beta\beta} & W_{\beta \overline{\beta}} \\
                          W_{\overline{\beta} \beta} & W_{\overline{\beta} \overline{\beta}} 
                         \end{array}
  \right) 
\left(
\begin{array}{c}
                          \beta_x \\
                          \overline{\beta}_x 
                         \end{array}
\right)
+ \frac{1}{2} \left(
\begin{array}{c}
                          1 \\
                          1 
                         \end{array}
\right)=0.
 \end{equation}
So for the solutions of the dDR system one has 
\begin{equation}
 \beta_x = -\frac{1}{2}\frac{1}{W_{\beta\beta}}.
\end{equation}
Thus regular sector of the dDR system (\ref{dLIA-Eqn}) is characterized by the condition \cite{KMM2}
\begin{equation}
 W_{\beta}=0, \qquad  W_{\beta \beta} \neq 0.  
\end{equation}
In  this sector $\beta$ and the derivatives $\beta_x$, $\beta_y$, i.e. curvature $K$, torsion $\tau$, and their derivatives are bounded. \par
Singular sector of dDR system is composed by solutions for which $W_{\beta \beta}=0 $. For these solutions $\beta$ (i.e. curvature and torsion) remain bounded while their derivatives explode. Such situation is usually referred as gradient catastrophe (see e.g. \cite{RY,Whi}). Generic gradient catastrophe is characterized by the conditions \cite{KMM2}  
\begin{equation}
\label{catcon-b}
 W_{\beta}=0, \qquad W_{\beta \beta}=0, \qquad W_{\beta \beta \beta} \neq 0. 
\end{equation}
 At $K \neq 0$ ($\beta \neq \overline{\beta}$) one also has $W_{\beta \overline{\beta}}=0$ and the conditions (\ref{catcon-b}), in terms of curvature and torsion are
\begin{equation}
\label{catcon-KT}
 W_K=0, \quad  W_\tau=0, \quad  W_{KK}=0,\quad W_{K \tau}=0, \quad  W_{KKK}\neq 0, \quad W_{\tau \tau \tau} \neq 0. 
\end{equation}
 Equations $W_{\tau \tau}=0$, $W_{K K \tau}=0$, $W_{K \tau \tau}=0$, are consequences of (\ref{catcon-KT}). \par
Last conditions (\ref{catcon-KT}) allow us to solve the conditions  $W_{KK}=W_{\tau \tau}=0$ with respect to $K$ and $\tau$. Substituting these expressions into the first two conditions (\ref{catcon-KT}), one gets two equations
\begin{equation}
 \label{cat-point}
f_1(x,y)=0, \qquad f_2(x,y)=0.
\end{equation}
Thus generic gradient catastrophe for the dDR system happens in a single point $(x_0,y_0)$ for given $W({\beta, \overline{\beta}})$, i.e. initial data for $K$ and $\tau$ \cite{KMM2}. For the focusing dNLS equation this fact has been first noted in \cite{DGK}. We would like to emphasize that in contrast to the hyperbolic case where the catastrophe locus is a curve in $x,y$ here the catastrophe happens in a point.
\section{Flutter of filament}
Gradient catastrophe for the dDR system means a special behavior of a filament at the point $x_0$ at time $y_0$.
  First one observes, using formula (20) in \cite{NSW} with $U=W=0$ and $V=K$, that the acceleration $\vec{a}$ for the velocity (\ref{LIA-X})
\begin{equation}
 \vec{a}=\vec{v}_t=\epsilon \vec{v}_y=\epsilon (K_y \vec{b} + K \vec{b}_y)= \epsilon\left[(-2K_x\tau-K \tau_x) \vec{b} - K_x \vec{t} + \tau^2 \vec{n}\right]
\end{equation}
explodes at $x_0$. Thus, at the moment $t_0=\epsilon y_0$  sharp bump is formed at the point $s_0=\epsilon x_0$ of the filament.\par
\begin{figure}[h!]
\centering
\begin{tabular}{ccccc}
\includegraphics[width=4cm, height=5cm]{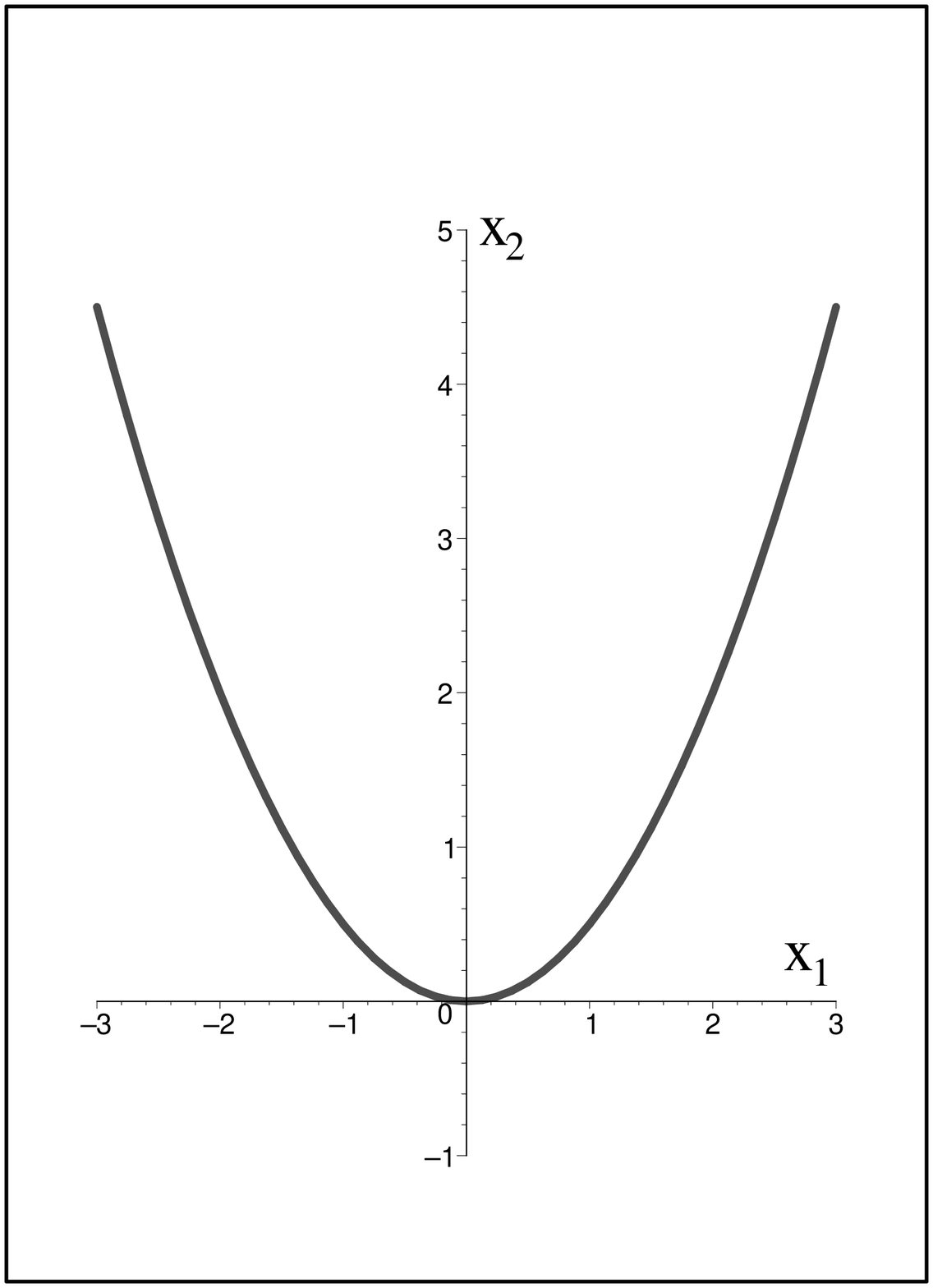}
& &
\includegraphics[width=4cm, height=5cm]{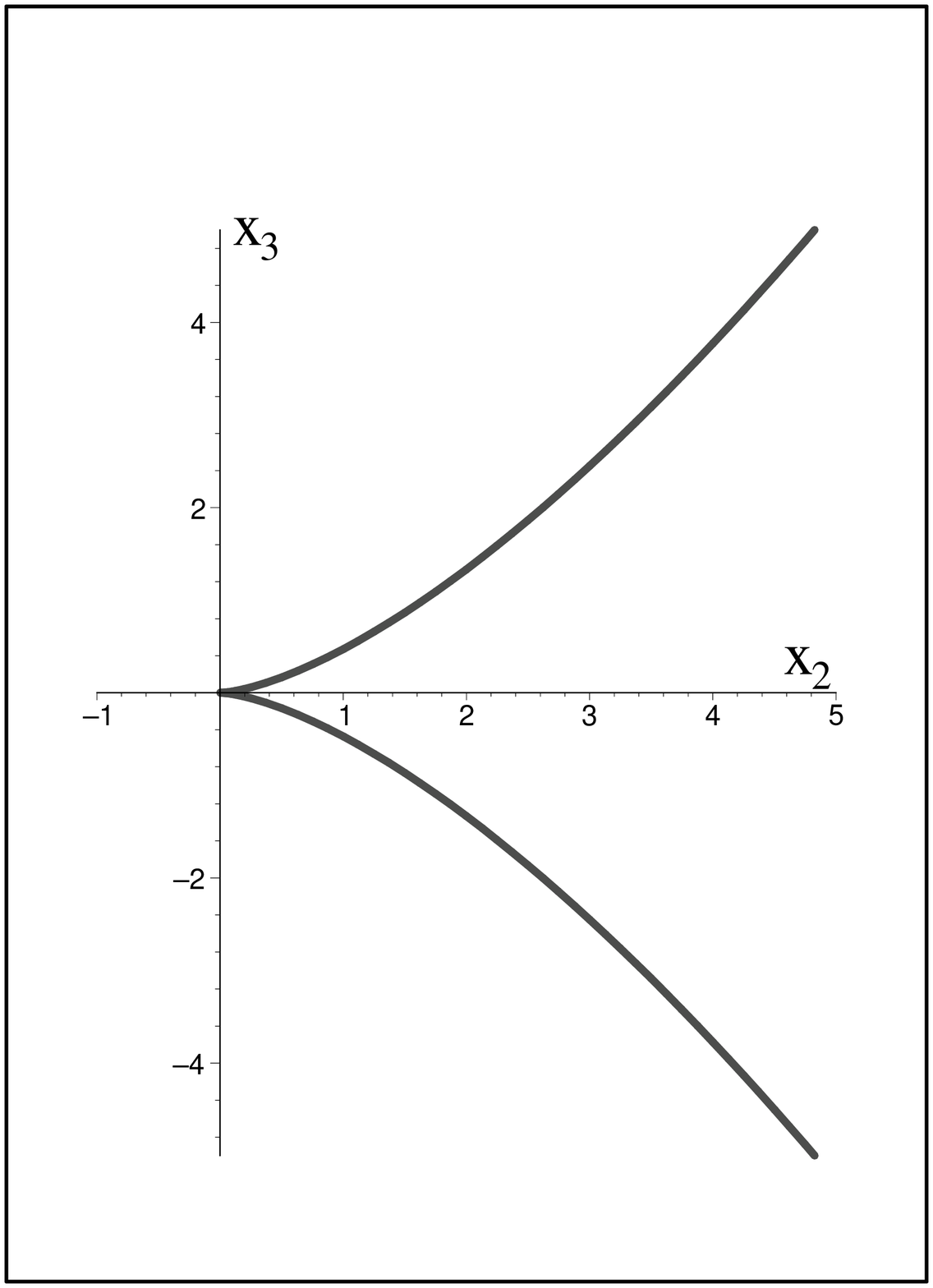}
& &
\includegraphics[width=4cm, height=5cm]{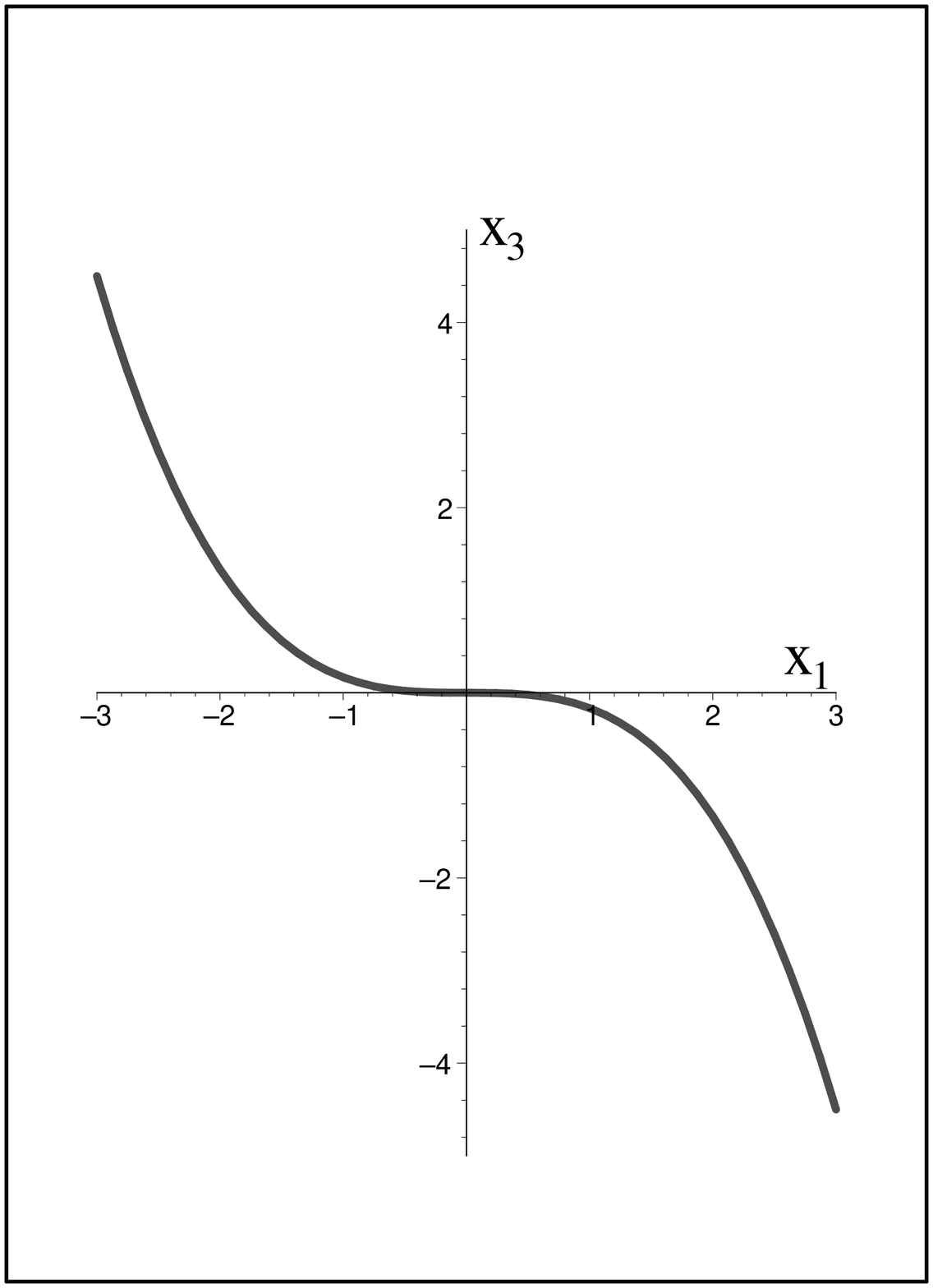}
\end{tabular}
\caption{The figure show a curve with $\tau>0$. At $\tau<0$ one should reflect the curve with respect to the axis $\vec{t}$ in the last figure.}
\label{fig-pcc}
\end{figure}

Local consideration reveals another peculiarity of behavior of filament around the point  $x_0$. It is well  known, that the coordinates of a curve in a neighborhood of a point in the reference system formed by the tangent vector $\vec{t}$, normal $\vec{n}$, and binormal $\vec{b}$ and origin in the point (see e.g. formula (6.7) \cite{Eis}) are
\begin{equation}
 \label{xxx}
\begin{split}
 & x_1=s-\frac{K^2}{6}s^3+\dots, \\
 & x_2=\frac{K}{2}s^2-\frac{K_s}{6}s^3+\dots, \\
 & x_3=\frac{K \tau}{6}s^3+\frac{1}{24}(2K_s\tau+K\tau_s)s^4+\dots \ .
\end{split}
\end{equation}
Passing to the slow variable $x$ and slow coordinates  $\tilde{x}_1=\epsilon x_1$, $\tilde{x}_2=\epsilon x_2$, $\tilde{x}_3=\epsilon x_3$, one gets the same series (\ref{xxx}) for slow variables with substitution $s \to x$. At ordinary point where $K$, $\tau$ and $K_x$, $\tau_x$ are bounded the first terms in r.h.s. of  (\ref{xxx}) dominate and the curve, locally, is a twisted cubic. Its projection on the osculating plane  (spanned by $\vec{t}$ and $\vec{n}$), normal plane  (spanned by $\vec{b}$ and $\vec{n}$) and rectifying plane (spanned by $\vec{t}$ and $\vec{b}$) are parabola ($x_2-\frac{K}{2}x_1^2=0$), cusp ($9Kx_3^2-2\tau^2 x_2^3=0$) and cubic ($x_3+\frac{K\tau}{6}x_1^3=0$) respectively (fig. \ref{fig-pcc}, see e.g. \cite{Eis}).
Characteristic feature of a curve (and a filament) in an ordinary point (i.e., outside the points of gradient catastrophe) is that it lies always on one side of positive direction of the normal and on one side of osculating plane  (depending on sign of torsion $\tau$). \par
\begin{figure}%[h!]
\centering
\includegraphics[width=5cm, height=7cm]{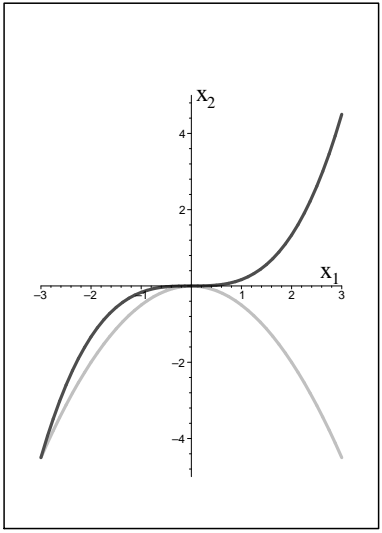}
\caption{Possible different forms of the projection on the osculating plane of a vortex filament near to the catastrophe point.}
\label{fig-clc}
\end{figure}

At the point $x_0$ of gradient catastrophe, the behaviors of a curve  changes drastically. Indeed, when $K_x$ and $\tau_x$ become large, the second terms in $\tilde{x}_2$
and $\tilde{x}_3$ become relevant. So, parabola in the osculating plane may
change sign or even convert into cubic curve (fig. \ref{fig-clc}) while in the
normal plane it could be a plane $(3,4)$ curve and so on. 
So, around the point $x_0$ of gradient catastrophe a filament oscillates from one side of the rectifying 
%and osculating
 planes to another and back. Such oscillation is quite similar to that of airfoil (see e.g. \cite{MilT}) and one can refer to these oscillation of a filament in the point of gradient catastrophe as a flutter.\par 
At each point of a curve there is a sphere which has contact of a third order at this point. It is called the osculating sphere (see e.g. \cite{Eis}). It has radius 
\begin{equation}
\label{R-osc} 
R^2=\frac{K^2\tau^2+K_s^2}{K^4\tau^2}.
\end{equation}
 and its center has coordinates
\begin{equation}
\label{X-osc}
 \vec{X}_o=\vec{X}+\frac{1}{K} \vec{n}-\frac{K_s}{K^2\tau}\vec{b}
\end{equation}
where $\vec{X}$ is the position vector of a point on the curve. \par
At the point of gradient catastrophe the radius of the osculating sphere and its center blow up to infinity.\par
Oscillation of a curve around the rectifying plane and blow up of the osculating sphere are geometrical features of the gradient catastrophe and flutter of a filament.
%%%%%%%%%%%%%%%%%%%%%%%%%%%%%%%%%%%%%%%%%%%%%%%%%%%%%%%%%%%%%%%%%%%%%%%%%%%%%%%%%%%%%%%%%%%%%%%%%%%%%%%%%%%%%%%%%%%%%%%
At the point $x_0,t_0$ of gradient catastrophe derivatives $K_x,\tau_x$  are so large that the terms of higher order derivatives in the DR equation (\ref{LIA-Eqn-e}) become relevant for finite $\epsilon$. Solutions of the full DR system (\ref{LIA-Eqn-e}) at small $\epsilon$ are almost indistinguishable from those of the dDR system at $t<t_0=\epsilon y_0$ (for dNLS equation see \cite{MK}--\cite{BT}). At $t \geq t_0$ the behavior of solutions of dDR and DR system are completely different. Solutions of the DR system at $t \geq t_0$ develop a zone of rapid oscillations which expand around the point $x_0$. For the equivalent dNLS equation this fact has been observed both analytically and numerically (see e.g. \cite{MK}--\cite{BT}).\par
An example of such a behaviors is given by the figure 13 of paper \cite{CMM}. In the terms of $K$ and $\tau$ the initial data from \cite{CMM} are $K(x,0)=2e^{-x^2}$ and $\tau(x,0)=-\tanh(x)$. Gradient catastrophe happens in $x_0=0$ and $y_0\simeq .25$. At this point the corresponding solution of DR system becomes oscillate.  As time $y$ growth a zone of rapid oscillations expands as $\sqrt{y-y_0}$ symmetrically around the point $x_0$. For the filament dynamics this effect is seen as the creation of rapid oscillations at the point of gradient catastrophe and  their subsequent expansion along the growing in time piece of filament (see fig. 3). 
\begin{figure}%[h!]
\centering
\includegraphics[width=12cm, height=5cm]{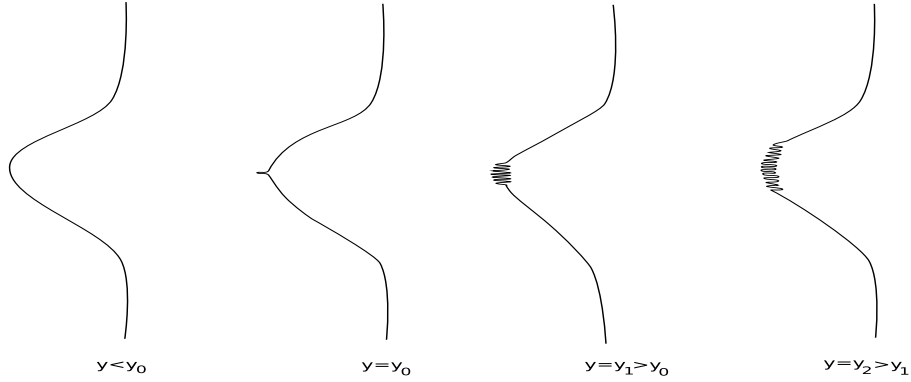}
%\caption{The figure show a curve with $\tau>0$. At $\tau<0$ one should reflect the curve with respect to the axis $\vec{t}$ in the last figure..}
\caption{Typical behavior of a flutter vortex line.}
\label{fig-dis}
\end{figure}
Such a behavior is an evidence of the flutter type instability of the vortex filament motion. 
\section{Elliptic umbilic catastrophe}
In order to understand better this phenomenon one should analyse in more detail the structure of the singularity and the regularizing mechanism for the gradient catastrophe. For the focusing dNLS equation such an analysis based on the $\epsilon$-expansion of the integrals of motions of the NLS/Toda equation has been performed in \cite{DGK}. Here we will follow a different approach discussed recently in \cite{KMA}.\par
Thus we consider a neighborhood of the gradient catastrophe point $x_0,y_0$ and denote the values of $\beta$ at this point by $\beta_0$. Following to the double scaling limit method (see e.g. \cite{DGZJ}--\cite{MAM}) we will look for solutions of the DR system near the point of gradient catastrophe  of the form
\begin{equation}
 \label{gc-point}
\begin{split}
x=x_0+\epsilon^{\alpha} x^* \\
y=y_0+\epsilon^{\sigma} y^* \\
\beta=\beta_0+\epsilon^{\gamma} \beta^* \\ 
\end{split}
\end{equation}
where $\epsilon \ll 1$ and numbers $\alpha,\sigma,\gamma$ should be fixed by further consideration. For the sake of simplicity we restrict ourselves to the case $y^*=0$. We first consider the function $W(\beta,\overline{\beta})$ (\ref{W}). One has
\begin{equation}
\label{Wexp}
\begin{split}
 &W(x_0+\epsilon^{\alpha} x^*,y_0,\beta_0+\epsilon^{\alpha} \beta^*)= W^0+\frac{1}{2}(\beta_0+\overline{\beta}_0) x^* \\
&+ \epsilon^\gamma \left[   
 \left(\frac{1}{2} x_0 +\frac{1}{4}({3\beta_0 + \overline{\beta}_0}) +\widetilde{W}_{\beta}^0 \right) \beta^*+
 \left(\frac{1}{2} x_0 +\frac{1}{4}({\beta_0 + 3\overline{\beta}_0}) +\widetilde{W}_{\overline{\beta}}^0 \right) \overline{\beta}^*\right] +\frac{1}{2} \epsilon^{\alpha+\gamma}x^* ({\beta^* + 3\overline{\beta}^*}) \\
&+\frac{1}{2} \epsilon^{2\gamma} \left[ 
\left(\frac{3}{4}y_0+\widetilde{W}^0_{\beta \beta}\right) {\beta^*}^2+
\left(\frac{3}{4}y_0+\widetilde{W}^0_{\overline{\beta}\overline{\beta}}  \right) {\overline{\beta}^*}^2+
\left(\frac{1}{2}+2\widetilde{W}^0_{{\beta}\overline{\beta}}  \right){\overline{\beta}^*}\beta^*
\right] \\
&+\frac{1}{6} \epsilon^{3\gamma} \left[
\widetilde{W}^0_{\beta\beta\beta} {\beta^*}^3+
2\widetilde{W}^0_{\beta\beta \overline{\beta}} {\beta^*}^2{\overline{\beta}^*}+
2\widetilde{W}^0_{\beta \overline{\beta} \overline{\beta}} {\beta^*}{\overline{\beta}^*}^2+
\widetilde{W}^0_{\overline{\beta}\overline{\beta}\overline{\beta}} {\overline{\beta}^*}^3
\right]+\dots
\end{split}
\end{equation}
where $W^0=W\Big{\vert}_{\beta=\beta_0}$, $\widetilde{W}^0_\beta = \frac{\partial \widetilde{W}}{ \partial \beta}
\Big \vert_{\beta=\beta_0}$, 
$\widetilde{W}^0_{\overline{\beta}} = \frac{\partial \widetilde{W}}{ \partial \overline{\beta}} \Big \vert_{\beta=\beta_0}$  and so on. \par
Hodograph equation (\ref{hodoW})  and conditions (\ref{catcon-b}) imply that
\begin{equation}
\label{consq-hodo}
 \frac{x_0}{2}+\frac{y_0}{4} (3\beta_0 +\overline{\beta}_0)+\widetilde{W}^0_\beta =0 , \qquad 
\frac{3}{4}y_0 + \widetilde{W}^0_{{\beta}{\beta}}=0.
\end{equation}
At curvature $K \neq 0$ the Euler-Poisson-Darboux equations (\ref{EPDe}) and its differential consequences
\begin{equation}
\label{consq-diff-EPD}
   2\beta W_{\beta \overline{\beta}} +2(\beta-\overline{\beta})W_{\beta \beta \overline{\beta}}
=W_{\beta\beta}-W_{\overline{\beta}\beta}, 
\end{equation}
imply that 
\begin{equation}
 W^0_{\beta \overline{\beta}}=0, \qquad \qquad 
 W^0_{\beta \beta \overline{\beta}}=
 W^0_{\beta \overline{\beta} \overline{\beta}}=0,
\end{equation}
i.e.
\begin{equation}
\label{W-cond-crit}
 \frac{y_0}{4}+\widetilde{W}^0_{\beta \overline{\beta}}=0, \qquad \qquad
 \widetilde{W}^0_{\beta \beta \overline{\beta}}=
 \widetilde{W}^0_{\beta \overline{\beta} \overline{\beta}}=0.
\end{equation}
Taking into account equations (\ref{consq-hodo}) and (\ref{W-cond-crit}), one finally gets
\begin{equation}
 W(x_0+\epsilon^{\alpha} x^*,y_0, \beta_0+\epsilon^\gamma \beta^*)=
W_0+\frac{1}{2}\epsilon^{\alpha} (\beta_0+\overline{\beta}_0)x^*+\frac{1}{2} \epsilon^{\alpha+\gamma} (\beta^*+\overline{\beta}^*)x^*+\frac{1}{3}\epsilon^{3\gamma} (a{\beta^*}^3+{\overline{a}\overline{\beta}^*}^3)+\dots 
\end{equation}
where $a=\frac{1}{2}\widetilde{W}^0_{\beta \beta \beta}$. Consequently near the point $x_0,y_0$ the hodograph equations take the form 
\begin{equation}
 W_{\beta^*}=\frac{1}{2}\epsilon^{\alpha+\gamma}x^*+\epsilon^{3\gamma}a{\beta^*}^2=0.
\end{equation}
 These equations readily imply that $\alpha=2\gamma$ and $\beta^*={x^*}^{1/2}$. So, near to the point  $x_0$
\begin{equation}
 \beta^*_{x^*} \sim (x-x_0)^{-{1/2}}. 
\end{equation}
Thus near to the point of gradient catastrophe the function $W$ is of the form
\begin{equation}
\label{W-x0y0}
 W=W_0+\frac{1}{2}\epsilon^{2\gamma}(\beta_0+\overline{\beta}_0)x^* +\epsilon^{3\gamma} W^*  
\end{equation}
where
\begin{equation}
 W^*=\frac{1}{2}x^* (\beta^*+\overline{\beta}^*)+\frac{1}{3}(a {\beta^*}^3+ \overline{a}{\overline{\beta}^*}^3)
\end{equation}
and $\beta=-\tau+iK$. Denoting $a^{1/3}\beta^*=U+iV$, one gets 
\begin{equation}
 W^*=x^*(f U + g V)+\frac{1}{3}U^3-UV^2
\end{equation}
where $f$ and $g$ are certain constants. In the Thom's catastrophe theory a function of this form is known as
 describing elliptic umbilic singularity (e.g. \cite{Thom,AGV}). So gradient catastrophe and flutter for vortex filament motion enter into the general scheme of Catastrophe Theory. For the focusing NLS equation the appearance of elliptic umbilic singularity has been observed for the first time in \cite{DGK}.
\section{Double scaling regularization and Painlev\'e-I equation}
 Next step is to regularize this singularity. Within the double scaling limit method adopted also in \cite{DGK} one begins with the full dispersive system. Performing appropriate double scaling limit one calculate the required regularizing terms. An approach discussed in \cite{KMA} suggests to proceed in opposite direction, namely, to begin with the original  (say dispersionless system), to modify the corresponding function $W$ adding to it the differential terms of lowest order with appropriate scaling and then to require that the critical point equations (\ref{Eul-Lag}) for $W$ are substituted by the Euler Lagrange equations for the modified $W_q$, i.e. 
\begin{equation}
 \frac{\delta W_q }{ \delta \beta}=0, \qquad 
 \frac{\delta W_q }{ \delta \overline{\beta}}=0. 
\end{equation}
Following this approach we first observe that the contributions of $\beta^*$ and $\overline{\beta}^*$ into the perturbation $W^*$ of $W$ are separated. It is quite natural to modify the function   (\ref{Wexp}) in such a way that this separation and reality property $W(\beta,\overline{\beta})=\overline{W(\beta,\overline{\beta})}$ are preserved. Thus, omitting inessential second term in (\ref{W-x0y0}), and choosing without lost of generality $\gamma=1$, one has the following modified $W$
\begin{equation}
 W_q=W_0+\epsilon^3\left[ \frac{1}{2}x^* (\beta^*+\overline{\beta}^*)+\frac{1}{3}(a {\beta^*}^3+ \overline{a}{\overline{\beta}^*}^3) \right] + \frac{1}{2}\epsilon^\delta 
\left[
b {\beta_{x^*}^*}^2+ \overline{b}\, {\overline{\beta}_{x^*}^*}^2
\right]
\end{equation}
   where $\delta$ and $b$ are appropriate constants. The corresponding Euler-Lagrange equation for $\beta^*$ and $\overline{\beta}^*$ is
\begin{equation}
 \frac{\partial W_q}{\partial \beta^*}  -\left( \frac{\partial W_q}{\partial \beta_{x^*}^*} \right)_{x^*}=0,
\end{equation}
 i.e.
\begin{equation}
 \epsilon^3\left[ \frac{1}{2}x^* \beta^*+a {\beta^*}^3 \right] -\epsilon^\delta 
\left[
b {\beta_{x^* x^*}^*}
\right]=0.
\end{equation}
To balance the first term one should choose $\delta=3$. \par
Thus, the modified $W_q$ is of the form
\begin{equation}
 W_q=W_0+\epsilon^3\left[ \frac{1}{2}x^* (\beta^*+\overline{\beta}^*)+\frac{1}{3}(a {\beta^*}^3+ \frac{1}{2} \overline{a}{\overline{\beta}^*}^3)+ \frac{1}{2} b {\beta_{x^*}^*}^2+ \frac{1}{2}\overline{b}\, {\overline{\beta}_{x^*}^*}^2\right]  \end{equation}
and the corresponding equation for $\beta^*$ is
\begin{equation}
\label{P-Iour}
 b\beta^*_{x^*x^*}=a{\beta^*}^2+\frac{1}{2}x^*.
\end{equation}
This equation is converted into the classical Painlev\'e-I equation (\ref{P-I}) by simple change of variables 
\begin{equation}
 x^*=\lambda \xi, \qquad \beta^*=-\frac{\lambda^3}{2b}\Omega
\end{equation}
with $a\lambda^5=-12b^2$.\par
So, the regularized behavior at the point of the gradient catastrophe of DR system, i.e. for the vortex filament dynamics is governed by the Painlev\'e-I equation. The relevance of this equation for the NLS/Toda system near to the critical point has been observed in a different way in \cite{DGK}. The correspondence between the solution of equation (\ref{P-Iour}) and those given by the formula (5.21) in \cite{DGK} is (at $b=1$) $u=K^2$, $v=2\tau$ and
\begin{equation}
 \begin{split}
  \overline{x}=&-i \varepsilon^{4/5} \left( r e^{i \psi}\right)^{-1/5} \left(-\frac{a}{36 K_0} \right)^{1/5}x^*,\\
  \overline{\beta}=&-ib \varepsilon^{2/5} \left( r e^{i \psi}\right)^{2} \left(-\frac{a}{36 K_0} \right)^{3/5}\beta^*
 \end{split}
\end{equation}
where $\overline{x}$, $\overline{\beta}$, $r$ and $\psi$ are defined in (2.19) and (2.22) of \cite{DGK}.
 One concludes also that $\varepsilon=\epsilon^{2/5}$.
In the paper (\cite{DGK}) it was conjectured that any generic solution of he NLS equation near to the critical point is described by the tritronque\'e solution of the Painlev\'e-I equation. Extension of this observation to our case allows to formulate the 
\begin{conj} 
 Generic behavior of the vortex filament near the point $x_0,y_0$ of gradient catastrophe is given by
\begin{equation}
\label{b-conj}
 \beta(s,t_0,\xi) \simeq \beta_0 
-\epsilon \left(-\frac{36}{a^3b} \right)^{1/5} \Omega_0\left(\left(-\frac{a}{12 b^3} \right)^{1/5} \frac{s-s_0}{\epsilon^3} \right)
\end{equation}
\end{conj}
where $\Omega_0$ is the tritronque\'e solution of the Painlev\'e-I equation (\ref{P-I}). \par
Properties of the solution (\ref{b-conj}) as well as other features of flutter for vortex filaments will be discussed elsewhere.\\
{\bf Acknowledgements} This work has been partially supported by PRIN grant no 28002K9KXZ and by FAR 2009 (\emph{Sistemi dinamici Integrabili e Interazioni fra campi e particelle}) of the University of Milano Bicocca.
%%%%%%%%%%%%%%%%%%%%%%%%%%%%%%%%%%%%%%%%%%%%%%%%%%%%%%%%%%%%%%%%%%%%%%%%%%%%%%%%%%%%%%%%%%%%%%%%%%%%%%%%%%%%%%%%%%%%%%%

\end{document}